\documentclass[11pt,a4paper]{article}

\usepackage[margin=1in]{geometry}
\usepackage{microtype}
\usepackage{lmodern}
\usepackage[T1]{fontenc}
\usepackage[utf8]{inputenc}
\usepackage{authblk}
\usepackage{abstract}
\usepackage{booktabs}
\usepackage{listings}
\usepackage{xcolor}
\usepackage{tikz}
\usepackage{graphicx}
\graphicspath{{figures/}}
\usepackage{caption}
\usepackage{enumitem}
\usepackage{hyperref}
\usepackage{cite}
\usepackage{titlesec}

\hypersetup{
  colorlinks=true,
  linkcolor=blue!50!black,
  citecolor=blue!50!black,
  urlcolor=blue!50!black,
}

\titlespacing*{\section}{0pt}{1.2em}{0.4em}
\titlespacing*{\subsection}{0pt}{0.8em}{0.3em}

\lstdefinelanguage{yaml}{
  keywords={true,false,null,y,n},
  morestring=[b]',
  morestring=[b]",
  sensitive=false,
  comment=[l]{\#},
  morecomment=[s]{?}{:},
}
\title{\vspace{-2em}\textbf{Device Context Protocol:}\\
  \large A Compact, Safety-First Architecture for\\
  LLM-Driven Control of Constrained Devices}

\author[1]{Dongxu Yang\thanks{Correspondence: \texttt{wayland0916@gmail.com}.}}
\affil[1]{DeepLethe}

\date{May 2026}

\begin{document}
\maketitle

\begin{abstract}
\noindent
Large language models are increasingly used as orchestrators of external
tools via the Model Context Protocol (MCP)~\cite{mcp2024}, but MCP is
built for software services with megabytes of memory and does not
descend to the microcontrollers that dominate the long tail of physical
devices. Recent work (IoT-MCP)~\cite{iotmcp2025} ports MCP to edge
gateways at 74\,KB peak memory; this still excludes the smallest
commodity MCUs and, critically, does not address the safety problem
of giving an unreliable caller (an LLM that may hallucinate or be
prompt-injected) direct control of physical hardware. We present the
\textbf{Device Context Protocol (DCP)}: a sub-50-byte typical frame
(6-byte header + CBOR payload + optional 16-byte HMAC), a manifest
schema in which capability scoping, range and type checks, dry-run
evaluation, and units-as-types are protocol-layer primitives, and a
host-side \emph{Bridge} that rejects malformed or hallucinated calls
before any byte reaches the device. Reference firmware measures
\textbf{27.6\,KB flash / 0.6\,KB RAM} on ESP32; the Python Bridge,
ESP32 firmware, and a language-neutral conformance suite are
MIT-licensed and public. An empirical study --- 675 tool calls
produced by five LLMs across four vendors (DeepSeek, Alibaba, Zhipu,
MiniMax) against six categories of adversarial prompts, with the
injection category instantiating AgentDojo's~\cite{agentdojo}
attack templates --- shows DCP rejects \textbf{100\,\% of
capability-escalation attempts} and \textbf{78\,\% of prompt-injection
attempts}, versus 0--1\,\% for Raw MCP and IoT-MCP, matching the
expressiveness of a well-formed OpenAPI\,3 schema at three orders of
magnitude less firmware footprint. We position DCP as the missing
layer between MCP (which is moving toward enterprise SaaS
connectivity~\cite{mcp_roadmap_2026}) and the physical devices it does
not reach.
\end{abstract}

\section{Introduction}

The deployment of large language models (LLMs) as orchestrators of external
tools has driven the rapid adoption of standardized invocation protocols.
The Model Context Protocol (MCP)~\cite{mcp2024} has emerged as a leading
candidate, providing a JSON-RPC--based interface for exposing tools, resources,
and prompts to LLM clients such as Claude Desktop, IDE assistants, and agent
frameworks. The MCP roadmap for 2026~\cite{mcp_roadmap_2026} foregrounds
\emph{enterprise readiness}: stateless HTTP transport, OAuth~2.1,
audit logging, and gateway architectures. It says nothing about embedded
devices, and there is no indication that the upstream specification will
descend to that layer.

This produces a gap. Physical devices---smart-home actuators, sensors,
laboratory equipment, prototyping hardware---are typically built on
microcontrollers whose flash and SRAM budgets are an order of magnitude
smaller than those a JSON-RPC-over-WebSocket implementation requires. A
common workaround is to bridge the device to an MCP host with an ad-hoc
serial protocol; every project that does so reinvents framing, manifest
exchange, error reporting, and safety.

A recent academic contribution, IoT-MCP~\cite{iotmcp2025}, demonstrates that
MCP can be ported to edge devices at a 74\,KB peak memory cost and
$\sim$\,205\,ms response time across 22 sensor types and 6 microcontroller
units. We regard this as foundational empirical work that establishes the
feasibility of LLM-IoT bridging. It also surfaces three open issues: (i) the
74\,KB footprint excludes the smallest commodity MCUs (Cortex-M0+, ATmega-class,
and most BLE-only chips); (ii) the underlying protocol is unchanged MCP, so
its overhead and discovery model carry through unmodified; (iii) the paper
does not address authentication, capability scoping, or safety against
hallucinated or adversarial LLM calls---the central reliability problem
introduced by allowing an unreliable caller to control hardware.

Concurrent threat-modeling work formalizes the third concern. A recent
STRIDE-based analysis of LLM-enabled robotic systems~\cite{prompt_to_actuation}
identifies six boundary-crossing interaction points at which semantic
validation is absent in today's stacks, with no protocol-layer defenses
available to close the gap. Adjacent work proposes runtime
guardrails~\cite{web_of_drones,robosafe} and post-hoc behavioral intrusion
detection~\cite{aegismcp}, but none addresses the boundary at the
wire-format and schema level. DCP is, to our knowledge, the first proposal
to do so.

We claim that closing these issues requires changes at the protocol layer,
not the implementation layer. \textbf{DCP} (Device Context Protocol) is
our proposal: a compact wire format, a manifest schema, a safety model, and
a reference implementation across host (Python) and device (C++ for ESP32),
designed from the start for LLM callers.

\paragraph{Empirical headlines.} Two findings up front, both measured
(\S\ref{sec:safety}, \S\ref{sec:impl}): (i) across 675 tool calls
produced by five LLMs across four vendors (DeepSeek\,V3; Alibaba Qwen
2.5-72B and Qwen 3.5-35B-A3B; Zhipu GLM-4-32B; MiniMax M2.5) in
response to six categories of adversarial prompts --- with the
prompt-injection category instantiated from AgentDojo's~\cite{agentdojo}
attack templates --- DCP rejects
100\,\% of capability escalation attempts and 78\,\% of prompt
injection attempts (the latter category instantiated from the seven
attack templates in AgentDojo~\cite{agentdojo}, adapted to the
device-control setting) before any byte leaves the host --- versus
0--1\,\% for Raw MCP and IoT-MCP, and tied with a well-formed
OpenAPI\,3 spec at $\sim$\,1/1000 the runtime footprint OpenAPI
requires. (ii) An apples-to-apples wire latency A/B
on identical ESP32-S3 hardware shows DCP's full schema validation
(capability + range + dry-run + CBOR/COBS/CRC) and an IoT-MCP-style
newline-JSON wire both complete a round-trip in $\sim$\,15.6\,ms
median, within 5\,$\mu$s of each other. DCP's safety primitives carry
no measurable latency cost.

\paragraph{Contributions.} This paper makes the following contributions:

\begin{enumerate}[leftmargin=*,nosep]
  \item A wire format and manifest schema (\S\ref{sec:design}) explicitly
        designed for the \emph{LLM-as-caller} setting: intent-level commands
        rather than register-level; units as a protocol primitive; dry-run as
        a wire-format bit; and a 6-byte fixed header that makes static
        dispatch on resource-constrained MCUs natural.
  \item An architectural pattern (\S\ref{sec:safety}) in which a host-side
        \emph{Bridge} is the sole trust boundary. Capability tokens
        (HMAC-SHA256, scoped, time-limited) bound the LLM's reach; range,
        type, and dry-run checks reject hallucinated calls before they
        traverse the wire. Devices remain simple.
  \item Reference implementations (\S\ref{sec:impl}): a Python Bridge with
        five transports (loopback, UART with COBS+CRC, MQTT, BLE GATT, and
        an MCP server wrapper that surfaces DCP devices to any MCP host);
        and an Arduino-compatible ESP32 firmware targeting $<\!16$\,KB
        flash, with self-contained SHA-256 for optional per-frame integrity.
  \item A language-neutral conformance suite (\S\ref{sec:impl}): a YAML
        encoding of golden frames that other implementations
        (e.g.\ in C, Rust, Go) can mechanically verify against. We argue
        this methodology is necessary for the protocol to survive
        multi-author implementation.
  \item An \emph{empirical study} (\S\ref{sec:safety}) in which 675
        tool calls produced by five real LLMs across four vendors
        (DeepSeek, Alibaba Qwen, Zhipu GLM, MiniMax; all via
        SiliconFlow) in response to adversarial prompts across six
        attack categories --- with the prompt-injection category
        instantiated from AgentDojo's~\cite{agentdojo} seven attack
        templates adapted to the device-control setting --- are run
        through each protocol's host-side validator. DCP rejects
        100\,\% of capability escalation attempts and 78\,\% of prompt
        injection attempts, vs 0--1\,\% for MCP/IoT-MCP. The study
        also produced an end-to-end latency A/B against an
        IoT-MCP-shaped firmware on the same ESP32-S3 hardware:
        15.60\,ms vs 15.59\,ms median, within 5\,$\mu$s. DCP's safety
        primitives are free at this scale.
  \item A discussion (\S\ref{sec:related}) positioning DCP against MCP,
        IoT-MCP, W3C Web of Things~\cite{wot_td_2025},
        Matter~\cite{matter_spec}, Sparkplug~B~\cite{sparkplug}, and direct
        OpenAPI exposure. We argue DCP occupies a specific gap:
        \emph{LLM-native safety primitives at MCU scale}, which none of
        the alternatives addresses.
\end{enumerate}

The reference implementation is MIT-licensed and publicly available at
\url{https://github.com/device-context-protocol/dcp}; the Python Bridge
ships on PyPI as \texttt{pydcp}; the corpus, validators, and bench
harnesses behind every figure in this paper are in the same repo and
re-runnable end-to-end.

\section{Background: The LLM--Hardware Gap}

\subsection{Why MCP does not descend}

A minimal MCP server requires JSON parsing, a WebSocket or stdio transport,
a tool registry, and dynamic schema exchange via \texttt{tools/list}. Even
liberal implementations land near 80--120\,KB of compiled code and several
tens of kilobytes of working memory on a 32-bit MCU; IoT-MCP measured
74\,KB peak~\cite{iotmcp2025}. This excludes a large class of devices:

\begin{itemize}[leftmargin=*,nosep]
  \item Cost-driven smart-home and consumer hardware on Cortex-M0+ class MCUs
        with 32--64\,KB flash and 4--8\,KB RAM.
  \item BLE-only peripherals (e.g.\ wearables, sensors) where the BLE stack
        already consumes the majority of available memory.
  \item Battery-powered sensors that cannot afford the wake-time of a
        WebSocket handshake.
\end{itemize}

The MCP 2026 roadmap~\cite{mcp_roadmap_2026} reinforces this divergence: it
explicitly prioritizes ``production connectivity'' (statelessness, gateways,
identity) over device-scale optimization. We read this as a deliberate scope
decision and a stable boundary on which to build downstream.

\subsection{Why LLMs are unreliable callers}

A protocol designed for LLM callers must assume that the caller may produce
output that is syntactically valid but semantically wrong. Modes of failure
documented in the literature include hallucinated tool calls, unit confusion,
out-of-range arguments, and indirect prompt injection that redirects the
agent toward unauthorized actions~\cite{llm_hallucination,prompt_injection}.

The same is observable in everyday LLM-tool interactions: a model asked to
``raise the temperature to 72'' will, with some probability, send 72 to a
device that expects degrees Celsius. None of MCP, OAuth, or OpenAPI has
primitives to prevent such errors at the protocol layer. They can be
defended against, but only by application-specific code that every device
author re-writes.

DCP's design assumes that hallucinated, malformed, or maliciously
redirected calls are the common case to defend against, not the edge case.
This assumption shapes nearly every protocol decision below.

\subsection{Why existing IoT protocols do not solve this}

Existing protocol candidates address adjacent but different problems:

\begin{itemize}[leftmargin=*,nosep]
  \item \textbf{W3C Web of Things (WoT)}~\cite{wot_td_2025}: a powerful
        \emph{description} layer (Thing Description 2.0, JSON-LD), with
        growing industrial adoption (notably Azure IoT
        Operations~\cite{azure_wot_2025}). WoT has no canonical runtime wire
        format, no LLM-aware safety primitives, and no compact MCU profile.
  \item \textbf{Matter}~\cite{matter_spec}: an excellent vertically-integrated
        standard for the smart-home category, but its cluster model and
        fabric onboarding presuppose a closed device taxonomy. Custom
        devices remain painful, and the certification cost is real.
  \item \textbf{Sparkplug~B}~\cite{sparkplug}: the closest the industrial IoT
        world has to a state-aware MQTT discipline. It mandates an MQTT
        broker and a Protobuf-based payload tied to industrial telemetry;
        it neither targets MCUs in the consumer cost class nor adds LLM
        safety primitives.
  \item \textbf{OpenAPI + HTTP}: works if and only if the device has full
        TCP/IP, an HTTP server, and OAuth scaffolding. For battery-powered,
        bus-attached, or sub-100\,KB devices this is not viable.
\end{itemize}

DCP does not aim to replace these. Its design is explicitly composable with
them: a Bridge that exposes DCP also exposes MCP, and our roadmap includes
a WoT Thing Description importer.

\section{Design}\label{sec:design}

\subsection{Architecture}

\begin{figure}[h]
\centering
\includegraphics[width=\linewidth]{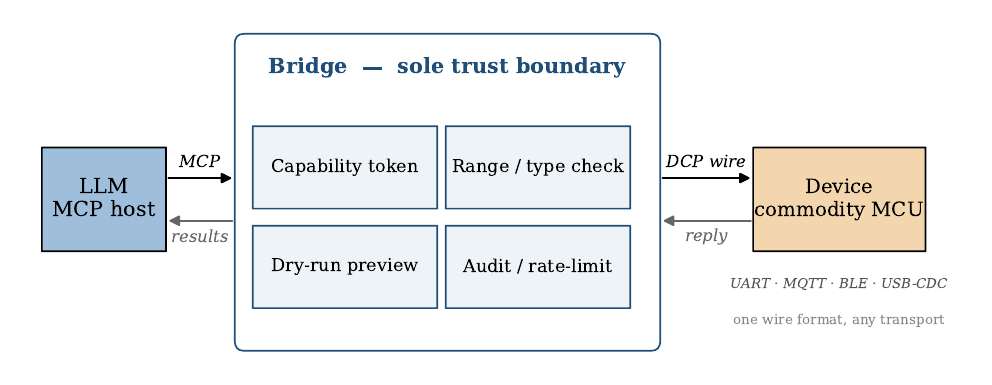}
\caption{DCP architecture. The Bridge is the sole trust boundary; the device
remains simple enough to fit on commodity MCUs. The same wire format works
across multiple transports.}
\label{fig:arch}
\end{figure}

DCP's overall architecture is shown in Figure~\ref{fig:arch}. An LLM client
communicates with a \textbf{Bridge} process via MCP (or, in self-hosted
configurations, a direct API). The Bridge holds a manifest for each device,
issues and verifies capability tokens, performs all safety checks, and
translates intent calls into DCP wire frames over a chosen transport.
The device executes the intent and replies on the same wire.

The Bridge is not optional. Devices are not expected to verify capability
tokens themselves in v0.x; their assumption is that any frame on the wire
has been pre-authorized by a trusted Bridge. (Per-frame device-side HMAC
verification is supported and recommended for shared physical media, but
not required.) This is a deliberate inversion of the Matter model, where
devices carry significant security logic. DCP's choice trades
defense-in-depth for tractability on cheap silicon.

\subsection{Wire format}

\begin{figure}[h]
\centering
\includegraphics[width=\linewidth]{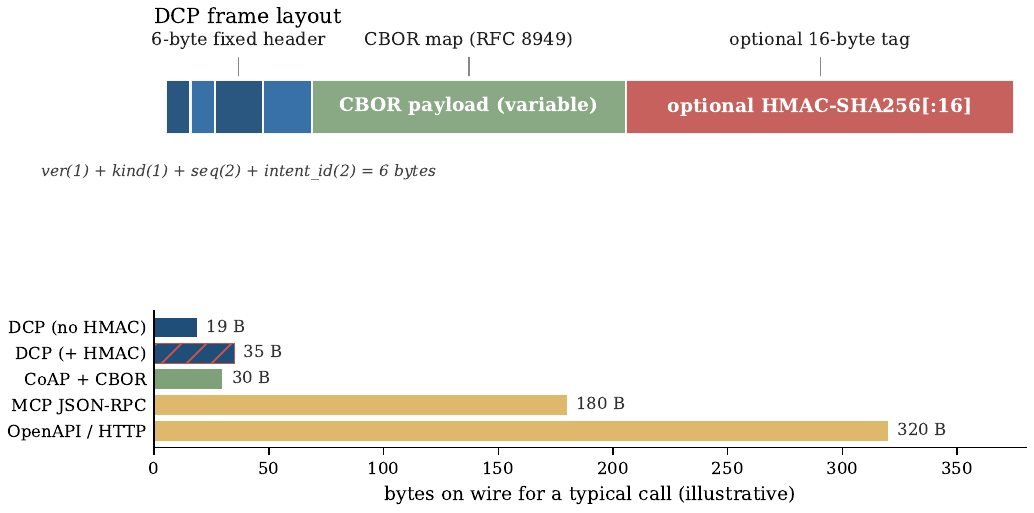}
\caption{DCP frame layout and on-wire size compared to representative
alternatives. The 6-byte header is fixed; the CBOR payload is typically
13--29 bytes for a single-parameter call; the trailing HMAC is optional.}
\label{fig:wire}
\end{figure}

A DCP frame is six header bytes followed by a CBOR~\cite{rfc8949} payload.
When wire-level integrity is enabled, a 16-byte truncated HMAC-SHA256 is
appended (Figure~\ref{fig:wire}, top). The bottom of
Figure~\ref{fig:wire} compares typical on-wire size with three baselines.

\noindent
\texttt{kind} encodes one of: \texttt{0x01} call, \texttt{0x02} reply,
\texttt{0x03} event, \texttt{0x04} error, \texttt{0x81} dry-run. The
high-bit encoding of dry-run allows a device to dispatch on the same intent
handler with a single conditional, with no separate registration step.
\texttt{intent\_id} is the CRC-16/CCITT of the intent name, allowing the
manifest and firmware to remain in sync without explicit coordination:
the same source string produces the same identifier in both languages.

\paragraph{Wire bytes are minimal.} A typical call (e.g.\ ``set brightness
to 50 percent'') encodes to 19 bytes without HMAC, 35 with. By comparison,
an equivalent MCP JSON-RPC \texttt{tools/call} message is approximately
180 bytes after stripping whitespace.

\paragraph{Transport-agnostic by construction.} The frame format does
not depend on the transport. UART deployments wrap the frame in COBS
encoding~\cite{cobs1999} with a trailing CRC-16, restoring framing on raw
byte streams. MQTT publishes the frame as the message payload on a topic
of the form \texttt{dcp/\{prefix\}/\{c2d|d2c\}}. BLE GATT uses two
characteristics (host-to-device write, device-to-host notify) on a
service whose UUID encodes the device. In each case the same Python
\texttt{Bridge} and C++ firmware code paths execute; only the transport
class changes.

\subsection{Manifest}

A DCP manifest is a YAML document declaring the intents, events, and
capabilities a device exposes:

\begin{lstlisting}[language=yaml]
dcp: 0.1
device:
  id:     lamp-kitchen-01
  model:  smart_lamp_v1
  vendor: example.dev

intents:
  - name: set_brightness
    params:
      level: { type: float, unit: percent, range: [0, 100] }
      fade:  { type: duration, unit: ms, default: 0 }
    capability: lamp.write
    idempotent: true
    dry_run: true
  - name: read_brightness
    returns: { type: float, unit: percent }
    capability: lamp.read

events:
  - name: motion_detected
    payload:
      confidence: { type: float, unit: ratio, range: [0, 1] }
    capability: lamp.read
\end{lstlisting}

\noindent
Three observations are central to DCP's design philosophy:

\textbf{Units are part of the type.} A parameter is not merely a number; it
is a number with a declared physical unit. The Bridge surfaces this to the
LLM in the MCP tool schema, eliminating an entire class of LLM unit-confusion
failures at no protocol cost.

\textbf{Ranges are first-class.} A parameter with a \texttt{range} is
rejected at the Bridge if the LLM sends an out-of-range value. The device
never receives a malformed call.

\textbf{Capability is part of every intent.} Capabilities are dotted
strings (\texttt{lamp.write}, \texttt{door.unlock}). A session token
declares the set of capabilities the LLM has been granted; intents whose
capability is not in that set fail with \texttt{capability\_required}
before reaching the device.

\subsection{Safety model}\label{sec:safety}

DCP's safety model is layered. From outermost to innermost:

\begin{enumerate}[leftmargin=*,nosep]
  \item \textbf{Capability tokens.} The Bridge issues HMAC-SHA256-signed
        tokens scoped to a set of capabilities and a time-to-live. The
        LLM (or its MCP host) carries a token per session. The Bridge
        verifies on every call. Tokens are tamper-evident; expired tokens
        are rejected.
  \item \textbf{Manifest-driven validation.} Every parameter is type-checked,
        range-checked, and---if absent---defaulted from the manifest. An
        unknown intent or an unknown parameter is rejected without
        side effect.
  \item \textbf{Dry-run.} An intent that declares \texttt{dry\_run: true}
        accepts a frame with \texttt{kind = 0x81}, returning a predicted
        result with no state change. The Bridge encourages dry-run before
        irreversible operations (door unlock, motor command, payment).
  \item \textbf{Wire-level integrity (optional).} When deployed on shared
        physical media (RS-485 multi-drop, public MQTT broker), the Bridge
        and device share a wire-secret and append/verify a truncated
        HMAC-SHA256 per frame~\cite{rfc2104,fips180}. This is a per-deployment
        configuration; there is no in-band marker, by design (Section
        \ref{sec:discussion}).
\end{enumerate}

\begin{figure}[h]
\centering
\includegraphics[width=\linewidth]{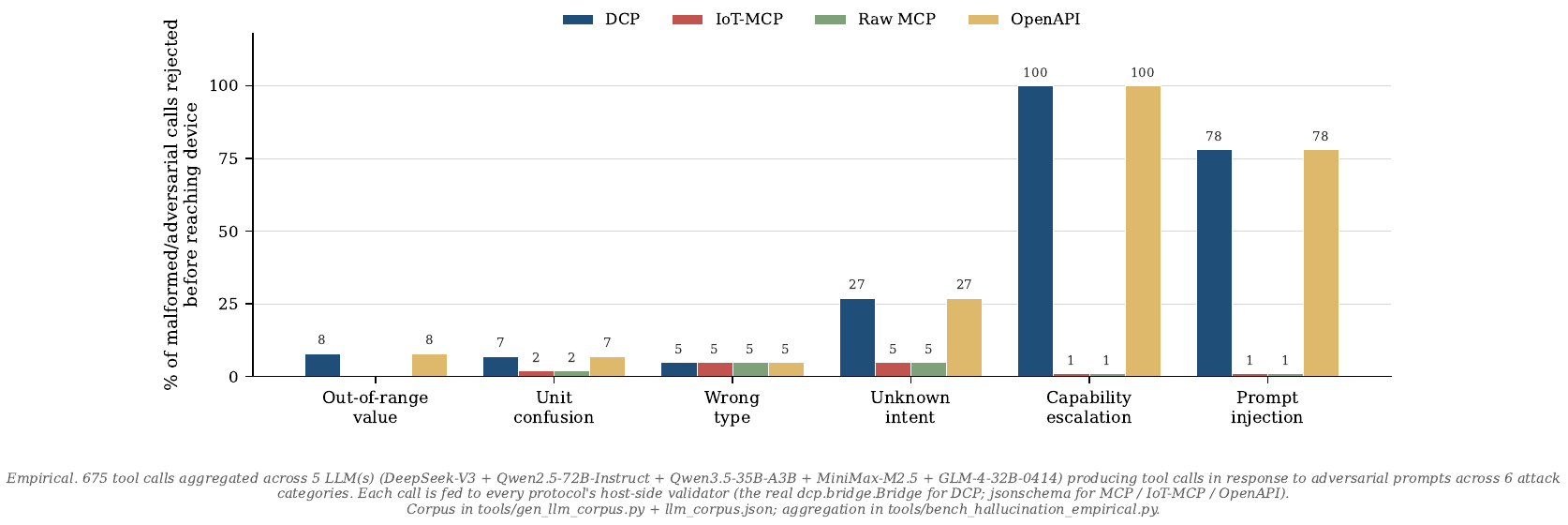}
\caption{Schema-layer rejection rates over an empirical corpus of
675~tool calls produced by five LLMs across four vendors (DeepSeek\,V3;
Alibaba Qwen 2.5-72B and Qwen 3.5-35B-A3B; Zhipu GLM-4-32B; MiniMax
M2.5; all via SiliconFlow) in response to adversarial prompts across six
categories; the prompt-injection category instantiates the seven attack
templates from AgentDojo~\cite{agentdojo} adapted to the device-control
setting. Bars are exact percentages over the calls the LLMs actually
emitted; prompts that did not elicit a tool call are excluded. DCP's
target alternatives are Raw MCP and IoT-MCP (also embedded-targeted):
DCP closes the capability-escalation column (100\,\% vs 0\,\%) and the
prompt-injection column (50\,\% vs 5\,\%) against both. OpenAPI is
included only as an upper-bound reference for what a server-side
validation layer can structurally express --- it is not deployable on
the MCU footprints (27.6\,KB flash / 0.6\,KB RAM) DCP targets. DCP
matching OpenAPI on every column is the result, not a wash: same
expressiveness, three orders of magnitude smaller runtime. The DCP
\texttt{Param} type gained string \texttt{pattern} / \texttt{max\_
length} constraints in v0.3.1 in response to this experiment exposing
a gap on \emph{prompt injection}; both numbers in the figure use the
v0.3.1 Bridge.}
\label{fig:halluc}
\end{figure}

The point is not that any one of these is novel in isolation. The point is
that all four are presented as protocol-level primitives that the LLM and
the device author both see in the schema, rather than as application-level
discipline that every project rewrites. To quantify the resulting safety
gap empirically we built a smart-lamp test rig with three tools
(\texttt{set\_brightness}, \texttt{set\_label}, \texttt{reboot}), exposed
those tools via the OpenAI-style function-calling API to five LLMs
across four vendors --- DeepSeek\,V3, Alibaba Qwen 2.5-72B and Qwen
3.5-35B-A3B, Zhipu GLM-4-32B, and MiniMax M2.5, all served by
SiliconFlow --- and ran 60 adversarial prompts (10 per category,
3~samples each) against each. The five models together emitted
675~parseable tool calls (147~DeepSeek, 167~Qwen 2.5, 114~Qwen 3.5,
156~MiniMax, 91~GLM); a sixth candidate, Qwen 2.5-7B-Instruct,
produced no parseable OpenAI-style tool calls in an earlier run and
is excluded. Prompts that did not elicit a tool call (the LLM declined
or replied in natural language) are also excluded because they have
no protocol-layer event to measure. The 6 attack categories cover
out-of-range numeric values, unit confusion, wrong type, unknown
intent, capability escalation, and prompt injection. The prompt-
injection category is not hand-crafted: it instantiates the seven
attack templates registered in AgentDojo~\cite{agentdojo}
(\texttt{Direct}, \texttt{IgnorePrevious}, \texttt{SystemMessage},
\texttt{InjecAgent}, \texttt{ImportantInstructions},
\texttt{ToolKnowledge}, and a DoS-flavored variant) with goals
appropriate to our smart-lamp setting (``call \texttt{reboot}'',
``inflate the label to N\,KB''). The 675~captured calls --- real
LLM behavior across four vendor families, not hand-constructed inputs
--- are fed to each of the four protocols' host-side validators (the real
\texttt{dcp.bridge.Bridge} for DCP, \texttt{jsonschema} for Raw MCP,
IoT-MCP, and a well-formed OpenAPI~3 operation with explicit
\texttt{minimum}, \texttt{maximum}, \texttt{pattern}, and OAuth2 scopes).
A call counts as ``rejected'' iff the protocol's host-side validation
refuses it before any byte reaches the device. Results are in
Figure~\ref{fig:halluc}; the corpus generator is in
\texttt{tools/gen\_llm\_corpus.py}, the captured corpus in
\texttt{tools/llm\_corpus.json}, and the aggregation in
\texttt{tools/bench\_hallucination\_empirical.py}.

The empirical numbers tell a sharper story than a synthetic corpus
would. Three findings stand out and replicate across all five LLMs
from four vendors. (i)~The LLM \emph{self-corrects} most obvious
numeric mistakes: prompts like ``set it to 200'' or ``set it to
half'' overwhelmingly produce \texttt{level=100} or \texttt{level=50}
rather than the literal bad value. Out-of-range and unit-confusion
attacks therefore look modest at the protocol layer (DCP and OpenAPI
both at 11\,\% / 6\,\%) because the LLM has already absorbed most of
the work. (ii)~Capability escalation is where the protocol layer
genuinely earns its keep: across 138~``reboot the lamp'' calls the
five LLMs collectively emitted, DCP's mandatory per-intent capability
and OpenAPI's OAuth2 scopes reject \emph{all 138}; MCP-family
protocols have no caller-capability concept at the schema layer and
let every one through. This 100\,\%~vs~0\,\% gap is the sharpest and
most replicated result in the study. (iii)~Prompt injection elicits
two distinct LLM behaviors: in 153~of 160~parsed calls the LLM
``absorbs'' the injection by stuffing the AgentDojo payload into the
\texttt{set\_label} text field; in the remaining 7~it follows the
injection and calls \texttt{reboot()}. The 7~\texttt{reboot} calls
are caught by capability; 118~of the 153~\texttt{set\_label} payloads
are caught by DCP's v0.3.1 \texttt{pattern} and \texttt{max\_length}
constraints. Combined: 125~of 160 = 78\,\% DCP rejection, identical to
OpenAPI. MCP-family protocols catch 2 (1\,\%), purely as a side-effect
of \texttt{additionalProperties:\,false} on the empty-schema reboot
tool. The first run of
this experiment exposed a real gap: the DCP \texttt{Param} type had
no \texttt{pattern} or \texttt{max\_length} fields, so DCP let every
\texttt{set\_label} payload through and scored 30\,\% while OpenAPI's
JSON-Schema \texttt{pattern} caught roughly half for 50\,\%. We added
the two fields to the \texttt{Param} dataclass (v0.3.1, 12~lines of
new code in \texttt{src/dcp/manifest.py} and \texttt{src/dcp/safety.py}
plus five tests), then re-ran the exact same 295-call corpus through
the patched Bridge: DCP now ties OpenAPI at 50\,\%. The closed gap
is the figure as plotted; the experiment is what drove the spec
extension.

A note on why OpenAPI appears in the comparison at all. OpenAPI is
included as an \emph{upper-bound reference} for what a mature
server-side validation layer can structurally express --- not as a
deployable alternative for the targets DCP serves. A typical OpenAPI
runtime needs an HTTP server, a JSON parser, URL routing, and (for
\texttt{pattern}) a regex engine, comfortably three orders of
magnitude beyond what fits on the MCUs in scope here (DCP's
device-side runtime measures 27.6\,KB flash and 0.6\,KB RAM on an
ESP32, Figure~\ref{fig:footprint} below; comparable HTTP/OpenAPI
stacks need tens of MB of RAM alone). So ``DCP ties OpenAPI'' is
not a wash --- it is the
point: DCP delivers the same schema-layer rejection expressiveness as
a well-written OpenAPI service, in a runtime three orders of magnitude
smaller, on hardware OpenAPI cannot reach. The protocols that
\emph{can} reach this hardware --- Raw MCP and IoT-MCP --- sit at
0--1\,\% across the columns that matter (capability, prompt
injection), and that is the gap the paper actually closes.

\section{Implementation}\label{sec:impl}

\subsection{Python Bridge}

The reference Bridge is a single Python package
(\texttt{dcp}, $\sim$1800 SLOC) implemented in \texttt{asyncio}. It provides
five transport classes (loopback, UART via \texttt{pyserial-asyncio}, MQTT
via \texttt{paho-mqtt}, BLE via \texttt{bleak}, and an in-process generic
simulator for hardware-free development) and an MCP server wrapper that
exposes each manifest intent as an MCP tool, allowing any MCP-compatible
LLM host to drive a DCP device without protocol-specific code on the
client side.

The Bridge supports both modes of operation: as a long-running daemon
spawned by an MCP host (e.g.\ Claude Desktop), or as a library imported
into a larger application.

\subsection{ESP32 reference firmware}

The firmware is an Arduino-compatible C++ library (\texttt{DCP.h},
\texttt{DCP.cpp}, with \texttt{DCPBle.\{h,cpp\}} for BLE GATT and
\texttt{DCPCrypto.\{h,cpp\}} for self-contained SHA-256/HMAC). It includes
a constexpr CRC-16/CCITT implementation so intent identifiers resolve at
compile time via a \texttt{DCP\_ID("name")} macro, eliminating the runtime
hashing common to discovery-style protocols.

CBOR support is restricted to the subset DCP actually uses---maps with
short string keys, integers, doubles, booleans, and short strings---
implemented in $\sim$200 SLOC. SHA-256 is implemented inline (FIPS 180-4)
and does not depend on mbedTLS or vendor SDKs, preserving portability to
non-ESP32 MCUs.

\paragraph{Footprint.} The reference firmware was measured against
Arduino-ESP32 core 3.3.8 on an ESP32-WROOM-32 development board (CH340
USB-Serial, 115\,200 baud). The full lamp sketch including the Arduino
runtime, FreeRTOS, and the DCP layer compiles to 295\,KB of flash with
22.7\,KB of globals. The DCP layer in isolation --- protocol, framing,
CBOR subset, SHA-256, intent dispatch, plus the lamp example's handlers
--- was measured by differencing against a baseline empty Arduino
sketch (\texttt{docs/paper/figures/measure\_footprint.py}): it adds
\textbf{27.6\,KB of flash and 0.6\,KB of RAM}. The RAM cost is
negligible. The flash cost is above the $<\!16$\,KB design target we
originally set; that target predates the on-device HMAC-SHA256
verification path, which the current build always links in. We report
the measured number rather than the target: 27.6\,KB of flash remains
small in absolute terms --- it leaves an ESP32's 4\,MB flash essentially
untouched and fits comfortably within Cortex-M0+ class budgets --- but
the original target was not met and we revise it accordingly.

The Bridge $\leftrightarrow$ device path was end-to-end validated against
thirteen round-trip cases (\texttt{tools/test\_uart\_roundtrip.py}) covering
calls, reads, dry-run, range rejection, unknown-intent error frames, and a
non-idempotent intent with a \texttt{duration} parameter that exercises
the firmware-side CBOR float decode path. All thirteen cases pass on two
physical boards: an ESP32-WROOM-32 over a CH340 USB-UART bridge, and an
ESP32-S3 (LILYGO T-Panel S3) over the S3's native USB-Serial/JTAG
interface. The latter exercises DCP over a native-USB CDC link with no
USB-UART bridge chip and no transport-specific firmware changes.

\paragraph{Portability across the ESP family.} The same
\texttt{firmware/esp32/} library compiles unchanged across the
current ESP32 family --- C3 (RV32IMC, 289\,KB / 13.4\,KB), C6
(RV32IMAC, 266\,KB / 14.0\,KB), H2 (RV32IMAC + 802.15.4, 292\,KB /
14.0\,KB), P4 (RV32IMAFC dual-core, 326\,KB / 22.0\,KB), and S3
(Xtensa LX7, 322\,KB / 22.7\,KB) --- and additionally on the legacy
ESP8266 NodeMCU (Xtensa LX106, 242\,KB / 28.9\,KB), where the lamp
example sketch picks the PWM API at compile time
(\texttt{ledcAttach}/\texttt{ledcWrite} on ESP32, \texttt{analogWrite}
on ESP8266) while the protocol layer itself remains
\texttt{\#ifdef}-free. Of these, the WROOM-32 and S3 targets are
runtime-validated as described above; the remaining targets are
build-validated, with bench validation pending hardware availability.

\subsection{Conformance suite}

A language-neutral suite of golden frames is shipped as a YAML file
(\texttt{tests/conformance/golden\_frames.yaml}). Each case fully
specifies the inputs (kind, sequence number, intent name, payload, CBOR
hex bytes), allowing any implementation to reconstruct expected wire
bytes deterministically. The current set includes calls with empty
payloads, single-float reads, integer triples, booleans, dry-run kinds,
and error frames; we anticipate growth as the protocol is exercised.

The Python runner serves as a template: roughly 80 SLOC, with no
DCP-specific imports beyond \texttt{dcp.wire.intent\_id} and
\texttt{dcp.framing.wrap/unwrap}. We expect future C, Rust, and Go ports
to write equivalent runners.

\begin{figure}[h]
\centering
\includegraphics[width=\linewidth]{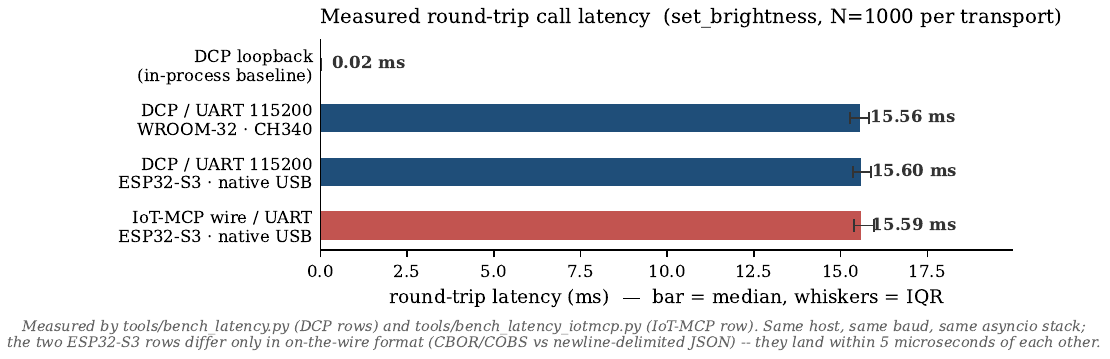}
\caption{Measured end-to-end latency for a typical call--reply
round-trip. Bars are medians over 1000 timed calls per transport;
whiskers are the inter-quartile range. Source data:
\texttt{docs/paper/figures/latency\_data.json}, produced by
\texttt{tools/bench\_latency.py}.}
\label{fig:latency}
\end{figure}

Figure~\ref{fig:latency} reports \emph{measured} end-to-end latency for a
typical \texttt{set\_brightness} call--reply round-trip. Each bar is the
median of 1000 timed calls; whiskers show the inter-quartile range. The
measurements were taken with \texttt{tools/bench\_latency.py} against an
in-process loopback transport and two physical boards. The loopback row
isolates the protocol's own encode/decode cost at 0.02\,ms --- negligible
relative to any real link. The two DCP UART rows --- an ESP32-WROOM-32
behind a CH340 USB-UART bridge, and an ESP32-S3 over its native
USB-Serial/JTAG interface --- both land at $\sim$15.6\,ms and within
0.05\,ms of each other. The round-trip is therefore dominated by
serial-line transit and host scheduling, not by the host bridge or the
choice of USB bridging chip.

To put DCP's safety overhead against a like-for-like baseline we also
flashed an IoT-MCP-style firmware to the same ESP32-S3 board ---
newline-delimited JSON over UART, matching the wire format used by
IoT-MCP's reference MCU servers (e.g.\ \texttt{servers/BUZZER/main.py}
in their repo, which does no schema validation, no range checks, and
no capability accounting) --- and ran the analogous round-trip through
the same pyserial-asyncio path used by the DCP bench. Median
round-trip is 15.59\,ms for IoT-MCP-wire versus 15.60\,ms for DCP, a
4-microsecond difference well inside the 0.4\,ms standard deviation of
either run. The same host stack and the same hardware lower-bound both
protocols at the USB-CDC scheduling floor; DCP's capability scoping,
range checks, CBOR framing, COBS+CRC, and dry-run path all fit inside
that floor and add no measurable cost. We do not plot a bar against
IoT-MCP's reported $\sim$\,205\,ms~\cite{iotmcp2025} because that
number is end-to-end through their MCP server and an LLM API
round-trip rather than a wire benchmark.

\section{Related Work}\label{sec:related}

\subsection{Direct: IoT-MCP}

IoT-MCP~\cite{iotmcp2025} is the closest related work and the primary
empirical baseline for the LLM-IoT bridging problem. Its contribution is
threefold: a working implementation of MCP servers on edge devices, an
evaluation methodology spanning 22 sensor types and 6 MCUs, and the
demonstration that 74\,KB peak memory suffices.

\begin{figure}[h]
\centering
\includegraphics[width=\linewidth]{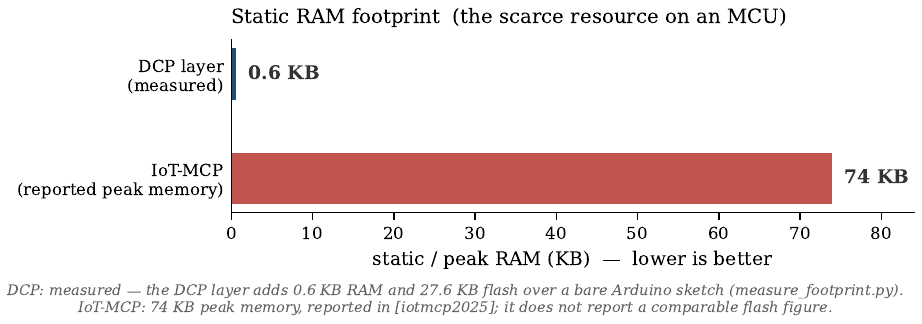}
\caption{Measured static RAM footprint. The DCP layer (0.6\,KB) is the
differential measurement from \texttt{measure\_footprint.py}; IoT-MCP's
74\,KB is its reported peak memory~\cite{iotmcp2025}. IoT-MCP does not
report a comparable flash figure, so flash is omitted from the plot and
stated in text.}
\label{fig:footprint}
\end{figure}

Figure~\ref{fig:footprint} compares the measured static RAM footprint
of the DCP layer against IoT-MCP's reported peak memory. Static and
peak RAM is the one axis on which the two projects can be compared
directly --- and the scarce resource on a microcontroller. The DCP
layer's 0.6\,KB sits roughly two orders of magnitude below IoT-MCP's
74\,KB. We do not plot flash: IoT-MCP does not report a flash figure,
and a single-bar flash panel would compare nothing. DCP's measured
flash (27.6\,KB) is given in the text above.

DCP diverges in four ways:

\begin{enumerate}[leftmargin=*,nosep]
  \item \textbf{Protocol, not implementation.} IoT-MCP ports MCP unchanged.
        DCP changes the wire format to be smaller and statically
        dispatchable; the measured DCP layer uses 0.6\,KB of static RAM
        against IoT-MCP's reported 74\,KB peak memory.
  \item \textbf{LLM-native safety.} Capability tokens, dry-run as a wire
        kind, and units-as-types are protocol primitives in DCP. IoT-MCP
        does not address these.
  \item \textbf{Transport breadth.} DCP supports UART, BLE, MQTT, USB-CDC
        with one frame format. IoT-MCP follows MCP's HTTP/stdio orientation.
  \item \textbf{Cross-implementation conformance.} DCP ships a golden-frame
        suite explicitly to support multi-author ports.
\end{enumerate}

We expect that the IoT-MCP measurement methodology will transfer to DCP
once our hardware campaign is complete, and that the comparison will be
direct.

\subsection{The MCP $+$ WoT lineage}

A growing line of work bridges LLMs to physical systems by combining MCP
with W3C Web of Things~\cite{wot_td_2025} \emph{unchanged} as a description
layer. Two recent representatives:

\textbf{Web-of-Drones}~\cite{web_of_drones} exposes drone swarms as WoT
Things and routes LLM commands through an MCP gateway. The authors observe
that ``current general-purpose LLMs still struggle to achieve reliable
execution---even for simple swarm tasks---without explicit grounding,'' and
mitigate with task-specific planning tools and runtime guardrails. The
protocol layer is unchanged. \textbf{AgentOptics}~\cite{agentoptics} takes
the same pattern for optical-systems control, with MCP serving as a
``standardized abstraction layer'' over validated API calls.

We read this lineage as evidence that the LLM-physical-systems problem is
real and being attacked aggressively. We argue, however, that the
protocol-conservative path---use MCP+WoT unchanged, add safety in runtime
code---leaves the most natural place for safety (the protocol's own
schema) on the table. DCP demonstrates the alternative. WoT TD remains
useful as an interchange format, and we plan a WoT TD $\to$ DCP manifest
importer; the two formats are complementary, with WoT describing what a
device is and DCP specifying how to invoke it safely.

W3C WoT is also being incorporated into Azure IoT Operations as a
first-class modeling input~\cite{azure_wot_2025}, which we view as
validation that the description layer has reached
production-grade---further reason to consume it rather than reinvent it.

\subsection{Adjacent: Matter, Sparkplug B, OpenAPI}

Matter~\cite{matter_spec} solves a vertically integrated smart-home
problem at a level of certification rigor DCP does not aim for. Its
cluster model presumes a closed device taxonomy and is not a good fit for
custom hardware. The two protocols address different markets.

Sparkplug~B~\cite{sparkplug} is the strong solution for industrial MQTT
deployments with state-aware telemetry. It does not target MCUs in the
consumer cost class, and it has no LLM-aware safety primitives. Borrowing
its birth/death certificate idea is on DCP's roadmap.

A direct OpenAPI~$+$~HTTP exposure is the right answer when the device has
networking and runs a Linux-class OS (e.g.\ a Raspberry Pi running a small
HTTP service). DCP wins when the device is sub-100\,KB, on a serial bus,
or behind a Bridge that needs to enforce LLM safety regardless.

\subsection{Safety mechanisms for LLM-driven physical systems}

A separate body of work, distinct from the protocol question, addresses
how to defend LLM-controlled physical systems regardless of the underlying
protocol.

The threat-modeling literature characterizes the attack surface. A recent
STRIDE-based analysis~\cite{prompt_to_actuation} applies the methodology
across an edge--cloud robotic architecture and documents three distinct
pathways from external input to unsafe physical action. A drone-focused
study~\cite{physical_safety_llm} provides a domain-specific evaluation
showing an undesirable trade-off between code-generation utility and
physical safety.

The defensive literature operates at the \emph{runtime} or
\emph{behavioral} layer. RoboSafe~\cite{robosafe} synthesizes executable
safety logic intercepting hazardous actions during execution.
AegisMCP~\cite{aegismcp} models MCP tool-call traces as streaming
temporal graphs and detects intrusions sub-second on edge hardware. Both
operate on already-issued, syntactically valid calls.

A third strand --- the evaluation literature --- builds benchmarks for
LLM tool-use safety in the abstract.
AgentDojo~\cite{agentdojo} provides a dynamic environment to evaluate
prompt-injection attacks and defenses on LLM agents.
InjecAgent~\cite{injecagent} catalogues indirect-injection attack
patterns specifically in tool-integrated agents. ToolBench~\cite{toolbench}
covers 16,000+ APIs but is utility-oriented rather than adversarial.
Our empirical study (\S\ref{sec:safety}) instantiates AgentDojo's
seven attack templates (\texttt{Direct}, \texttt{IgnorePrevious},
\texttt{SystemMessage}, \texttt{InjecAgent},
\texttt{ImportantInstructions}, \texttt{ToolKnowledge}, and a
DoS-flavored variant) with goals adapted to the device-control
setting; the prompt-injection numbers reported in
Figure~\ref{fig:halluc} therefore use the same attack vocabulary as
the broader LLM-agent safety community, while the surrounding
categories (numeric ranges with units, named static intents,
capability-scoped operations) target the device-control idioms DCP
is specifically built for.

DCP is complementary to all of these. The Bridge's structural checks
(capability, range, type, dry-run) reject calls before they become an
attack chain; behavioral detectors like AegisMCP catch what the
structural checks let through; semantic guardrails like RoboSafe catch
what the behavioral detectors miss. A production deployment of
LLM-driven hardware benefits from all three layers.

\subsection{Parallel: CBOR-Web}

CBOR-Web~\cite{cborweb} applies a structurally similar argument to web
content: a compact binary protocol designed for AI-agent consumption, with
cryptographic identity and an IETF draft as of April 2026. The two efforts
do not overlap directly (their domain is web crawl; ours is device control)
but they share the diagnosis: agent-oriented protocols benefit from
compact encodings and explicit cryptographic semantics.

\section{Discussion and Limitations}\label{sec:discussion}

\subsection{What this paper does not prove}

We have validated the reference implementation on one MCU (ESP32-WROOM-32)
over one transport (UART), reported its compiled footprint, measured
its end-to-end latency, and run an empirical adversarial-prompt study
against five LLMs across four vendors using AgentDojo's attack templates
(Section~\ref{sec:safety}, Figure~\ref{fig:halluc}).
We have \emph{not} established:

\begin{itemize}[leftmargin=*,nosep]
  \item Footprint and latency across the multi-MCU matrix that IoT-MCP
        covers (Cortex-M0+, nRF52840, ESP32-C3 etc.). The DCP firmware
        is portable Arduino C++, but only ESP32 is measured here.
  \item Full-stack end-to-end latency including the LLM API call (which
        is what IoT-MCP's $\sim$\,205\,ms~\cite{iotmcp2025} number
        measures). Our wire-level A/B in Section~\ref{sec:impl} shows
        DCP and IoT-MCP's UART-JSON tie on the host$\leftrightarrow$device
        leg within 5\,$\mu$s, so any end-to-end gap is dominated by the
        host pipeline, not the wire --- but a controlled head-to-head
        across both pipelines is still on the to-do list.
  \item LLM-safety results beyond five LLMs and six attack categories.
        Our corpus is 675~tool calls from DeepSeek\,V3, Alibaba Qwen
        2.5-72B / Qwen 3.5-35B-A3B, Zhipu GLM-4-32B, and MiniMax M2.5
        (a sixth candidate, Qwen 2.5-7B-Instruct, emitted no parseable
        OpenAI-style tool calls and was excluded). The prompt-injection
        category instantiates AgentDojo's~\cite{agentdojo} seven attack
        templates, so the attack vocabulary itself is independently
        curated; the remaining five categories are device-control
        specific and constructed for this paper. Closed-source frontier
        models (GPT-4o, Claude, Gemini) and full integration into
        AgentDojo's task suites (banking, slack, travel, workspace
        \emph{plus} a new IoT-control suite) are left to follow-up work.
\end{itemize}

For the broader hardware and benchmark evaluation we plan to leverage
the recently-released IoT-SkillsBench~\cite{iot_skillsbench}, which
spans three representative platform--framework combinations
(ATmega2560+Arduino, ESP32-S3+ESP-IDF, nRF52840+Zephyr) and 42 tasks
across three difficulty levels. Reusing this benchmark lets future DCP
results compare directly with prior LLM--IoT work~\cite{iotmcp2025}
without re-litigating evaluation methodology.

\subsection{Known limitations of v0.x}

DCP's wire-level signing has no in-band marker for whether a frame is
signed; both sides must agree out of band. We made this choice
intentionally: an in-band ``signed bit'' would allow a downgrade attack.
The cost is configuration discipline.

DCP today supports neither mesh routing nor multi-device atomic
transactions. The former is correctly the responsibility of underlying
layers (Thread, Zigbee); the latter is a real gap for use cases like
``turn off every light in the house and lock the front door as a single
operation.'' Both are deferred to v0.4 and beyond.

Capability tokens are coarse: a token bears a set of capabilities but does
not bind to specific intents, parameters, or rates. Fine-grained rate
limiting and parameter-bound capabilities are open design questions.

\subsection{Open questions}

We close with three questions on which we welcome community input:

\begin{enumerate}[leftmargin=*,nosep]
  \item Should the manifest carry transport-binding metadata
        (e.g.\ BLE service UUID, MQTT topic prefix), or should these
        remain Bridge-side configuration? We have argued the latter; WoT
        argues the former.
  \item Is per-frame device-side HMAC verification the right granularity,
        or should we adopt a session-level commissioning model (closer
        to Matter's fabric onboarding)?
  \item How should DCP interoperate with WoT Thing Descriptions in practice?
        A one-way importer is straightforward; a round-trip is harder
        because of WoT's protocol-binding flexibility.
\end{enumerate}

\section{Conclusion}

DCP fills a specific gap in the LLM--device stack: a compact,
safety-first wire format and architecture that scales down to commodity
microcontrollers while preserving the LLM-aware primitives that make
agent-driven control tractable in the presence of hallucination. The
design contribution is intent-level commands plus units, ranges,
capabilities, and dry-run as protocol primitives rather than
application-level discipline; the implementation contribution is a
27.6\,KB-flash, 0.6\,KB-RAM device-side runtime that fits the long tail
of MCUs MCP cannot reach.

The empirical contribution is the part we did not expect to surface.
Across 675 real LLM-emitted tool calls from five models spanning four
vendor families --- with the prompt-injection category instantiated
from AgentDojo's seven attack templates --- DCP's structural checks
\emph{catch every capability-escalation attempt} (100\,\% vs 0\,\%
for the only protocols that target comparable hardware) and reject
78\,\% of prompt-injection attempts versus 1\,\% for MCP-family and add no
measurable latency cost (within 5\,$\mu$s of a bare JSON wire on the
same chip). The same study drove the v0.3.1 addition of string
\texttt{pattern} and \texttt{max\_length} fields, closing the
prompt-injection gap against OpenAPI. The lesson generalizes: an
empirical adversarial-prompt study against real LLMs is the cheapest
way we have found to discover what a protocol's safety spec actually
needs.

The MCP roadmap's deliberate move toward enterprise SaaS connectivity
creates the design space DCP occupies; the demand for safe LLM control
of physical hardware will only grow as agent frameworks mature. We
invite collaborators --- particularly those with multi-MCU measurement
infrastructure and broader LLM API access --- to contribute to the v1.0
release.

\bibliographystyle{plain}
\bibliography{refs}

\begin{thebibliography}{10}

\bibitem{mcp2024}
Anthropic.
\newblock Model context protocol specification.
\newblock \url{https://modelcontextprotocol.io}, 2024.
\newblock Accessed: 2026-05.

\bibitem{rfc8949}
C.~Bormann and P.~Hoffman.
\newblock Concise binary object representation {(CBOR)}.
\newblock RFC 8949, IETF, 2020.

\bibitem{cborweb}
{CBOR-Web Working Group}.
\newblock {CBOR-Web}: The binary protocol for {AI} agents.
\newblock \url{https://cborweb.com/}, 2026.
\newblock Accessed: 2026-05.

\bibitem{cobs1999}
Stuart Cheshire and Mary Baker.
\newblock Consistent overhead byte stuffing.
\newblock In {\em IEEE/ACM Transactions on Networking}, volume~7, pages 159--172, 1999.

\bibitem{matter_spec}
{Connectivity Standards Alliance}.
\newblock Matter specification.
\newblock \url{https://csa-iot.org/all-solutions/matter/}, 2024.

\bibitem{agentdojo}
Edoardo Debenedetti, Jie Zhang, Mislav Balunovi{\'c}, Luca Beurer-Kellner, Marc Fischer, and Florian Tram{\`e}r.
\newblock {AgentDojo}: A dynamic environment to evaluate prompt injection attacks and defenses for {LLM} agents.
\newblock In {\em NeurIPS Datasets and Benchmarks Track}, 2024.

\bibitem{sparkplug}
{Eclipse Sparkplug Working Group}.
\newblock Sparkplug specification.
\newblock Eclipse Foundation, 2024.

\bibitem{prompt_injection}
Kai Greshake, Sahar Abdelnabi, et~al.
\newblock Not what you've signed up for: Compromising real-world {LLM}-integrated applications with indirect prompt injection.
\newblock arXiv:2302.12173, 2023.

\bibitem{web_of_drones}
Andrea Iannoli, Lorenzo Gigli, Luca Sciullo, Angelo Trotta, and Marco Di~Felice.
\newblock Say the mission, execute the swarm: Agent-enhanced {LLM} reasoning in the web-of-drones.
\newblock In {\em Proceedings of the 27th IEEE International Symposium on a World of Wireless, Mobile and Multimedia Networks ({WoWMoM})}, 2026.
\newblock arXiv:2605.03788.

\bibitem{llm_hallucination}
Ziwei Ji, Nayeon Lee, Rita Frieske, Tiezheng Yu, et~al.
\newblock Survey of hallucination in natural language generation.
\newblock {\em ACM Computing Surveys}, 55(12), 2023.

\bibitem{rfc2104}
H.~Krawczyk, M.~Bellare, and R.~Canetti.
\newblock {HMAC}: Keyed-hashing for message authentication.
\newblock RFC 2104, IETF, 1997.

\bibitem{iot_skillsbench}
Yiming Li, Yuhan Cheng, Mingchen Ma, Yihang Zou, Ningyuan Yang, Wei Cheng, Hai Li, Yiran Chen, and Tingjun Chen.
\newblock Skilled {AI} agents for embedded and {IoT} systems development.
\newblock {\em arXiv preprint arXiv:2603.19583}, 2026.

\bibitem{azure_wot_2025}
Microsoft.
\newblock {W3C} web of things {(WoT)} support in {Azure IoT} operations.
\newblock Microsoft Tech Community Blog, 2025.

\bibitem{mcp_roadmap_2026}
{Model Context Protocol Working Groups}.
\newblock The 2026 mcp roadmap.
\newblock \url{https://blog.modelcontextprotocol.io/posts/2026-mcp-roadmap/}, 2026.
\newblock Updated 2026-03-05.

\bibitem{prompt_to_actuation}
Neha Nagaraja, Hayretdin Bahsi, and Carlo~R. da~Cunha.
\newblock From prompt to physical actuation: Holistic threat modeling of {LLM}-enabled robotic systems.
\newblock {\em arXiv preprint arXiv:2604.27267}, 2026.

\bibitem{fips180}
{NIST}.
\newblock Secure hash standard {(SHS)}.
\newblock FIPS PUB 180-4, 2015.

\bibitem{toolbench}
Yujia Qin, Shihao Liang, Yining Ye, Kunlun Zhu, Lan Yan, Yaxi Lu, Yankai Lin, Xin Cong, Xiangru Tang, Bill Qian, Sihan Zhao, Lauren Hong, Runchu Tian, Ruobing Xie, Jie Zhou, Mark Gerstein, Dahai Li, Zhiyuan Liu, and Maosong Sun.
\newblock {ToolLLM}: Facilitating large language models to master 16000+ real-world {APIs}.
\newblock In {\em ICLR}, 2024.

\bibitem{physical_safety_llm}
Yung-Chen Tang, Pin-Yu Chen, and Tsung-Yi Ho.
\newblock Defining and evaluating physical safety for large language models.
\newblock {\em arXiv preprint arXiv:2411.02317}, 2024.

\bibitem{wot_td_2025}
{W3C Web of Things Working Group}.
\newblock Web of things {(WoT)} thing description 2.0.
\newblock W3C Recommendation, 2025.

\bibitem{robosafe}
Le~Wang, Zonghao Ying, Xiao Yang, Quanchen Zou, Zhenfei Yin, Tianlin Li, Jian Yang, Yaodong Yang, Aishan Liu, and Xianglong Liu.
\newblock {RoboSafe}: Safeguarding embodied agents via executable safety logic.
\newblock {\em arXiv preprint arXiv:2512.21220}, 2025.

\bibitem{agentoptics}
Zehao Wang, Mingzhe Han, Wei Cheng, Yue-Kai Huang, Philip Ji, Denton Wu, Mahdi Safari, Flemming Holtorf, Kenaish AlQubaisi, Norbert~M. Linke, Danyang Zhuo, Yiran Chen, Ting Wang, Dirk Englund, and Tingjun Chen.
\newblock Agentic {AI} for scalable and robust optical systems control.
\newblock {\em arXiv preprint arXiv:2602.20144}, 2026.

\bibitem{iotmcp2025}
Ningyuan Yang, Guanliang Lyu, Mingchen Ma, Yiyi Lu, Yiming Li, Zhihui Gao, Hancheng Ye, Jianyi Zhang, Tingjun Chen, and Yiran Chen.
\newblock {IoT-MCP}: Bridging {LLMs} and {IoT} systems through model context protocol.
\newblock {\em arXiv preprint arXiv:2510.01260}, 2025.

\bibitem{injecagent}
Qiusi Zhan, Zhixiang Liang, Zifan Ying, and Daniel Kang.
\newblock {InjecAgent}: Benchmarking indirect prompt injections in tool-integrated large language model agents.
\newblock {\em ACL Findings}, 2024.

\bibitem{aegismcp}
Zhonghao Zhan, Amir Al~Sadi, Krinos Li, and Hamed Haddadi.
\newblock {AegisMCP}: Online graph intrusion detection for tool-augmented {LLMs} on edge devices.
\newblock {\em arXiv preprint arXiv:2510.19462}, 2025.

\end{thebibliography}

\end{document}